\documentclass[twocolumn,amsmath,amssymb,prl,aps,showkeys]{revtex4-1}
\usepackage{graphicx}
\usepackage{epstopdf}
\usepackage{amsmath,amssymb}
\usepackage{mathrsfs}
\usepackage{color}
\usepackage{afterpage}
\usepackage[version=3]{mhchem}
\usepackage[normal]{subfigure}
\begin{document}
\title{Exploring semiconductor substrates for Silicene epitaxy}
\author{Amrita Bhattacharya$^\dagger$, Saswata Bhattacharya$^\dagger$, Gour P. Das$^*$}
\affiliation{Dept. of Materials Science, Indian Association for the Cultivation of Science, Kolkata-32, India}
\date{\today}
\begin{abstract}
We have carried out first-principles based DFT calculation on electronic properties of
 silicene monolayer on various (111) semi-conducting surfaces. We find that the relative stability and other properties of the silicene overlayer depends sensitively on whether the interacting top layer of the substrate is metal or non-metal terminated. The nature of silicene-monolayer on the metal terminated surface can be metallic or even magnetic, depending upon the choice of the substrate. The silicene overlayer undergoes n-type doping on
 metal terminated surface while it undergoes p-type doping on non metal terminated surfaces of the semiconductor substrates.
\end{abstract}
\maketitle
From first principles
calculations, the 2D mono-layers of Silicon, known as silicene, has also been predicted to be
stable\cite{r1,r2,r3} with graphene like
semi-metallic characteristics viz. a linear dispersion relation at
the K point\cite{r4,r5,r6}. The epitaxial growth of Ag thin film on Si(111) substrate has been
explored by experimentalists in the past. The limited intermixing between Ag thin
film and Si(111) substrate, make the choice of substrate ideal, for
the growth of Ag thin films, so, it was natural to conjecture a reverse case. Infact, the experimental evidence of formation of silicene on 
Ag(001), Ag(110), Ag(111) substrates has been reported by various groups \cite{r7,r8,r9,r10,r11,r12}. The formation of silicene on Ag(110) substrate has also been studied using first principles
methods by A. Kara \emph{et al} \cite{r13,r14}. More recently, evidence of epitaxial Silicene on non silver ZrB$_2$-0001 substrate \cite{r15} and Ir(111) substrate \cite{r16} has also been reported. Moreover, silicene monolayer on Ag(111) was found to undergo a phase
transition to two types of mirror-symmetric boundary-separated rhombic phases at temperatures below
40 K by scanning tunneling microscopy \cite{r17}. However, the possibility of existence of other suitable substrates for formation of silicene monolayer can not be ruled out and has not been exhaustively explored till date.\\
In this communication, we report from our first principles based calculations, the
bonding, stability and electronic structure of silicene monolayer,
when epitaxially grown on various Group II-VI and Group III-V
semiconductor substrates viz.
AlAs(111), AlP(111), GaAs(111), GaP(111), ZnS(111) and ZnSe(111). 
The nature of silicene monolayer on the metal
terminated surface of these substrates can be metallic or even magnetic, 
depending upon the choice of the substrate. The silicene overlayer undergoes
n-type doping on metal terminated surface while it undergoes p-type doping on
non metal terminated surfaces of the semiconductor substrates. \\
Our calculations have been carried out using first-principles
density functional theory (DFT)\cite{r23,r24} based total energy calculations. We have used
VASP \cite{r25} code with projected augmented wave
(PAW) potential\cite{r26} and Perdew-Burke-Ernzerhof (PBE) exchange correlation
functional\cite{r27} within generalized gradient
approximation (GGA). An energy cut off of 600 eV has been used. The
\underline{k}-mesh was generated by Monkhorst-Pack
method and the results were tested for convergence with respect to
mesh size [8$\times$8$\times$1]. However, for the generation of the 
electronic density of states (DOS) and band
structure plots, higher values of \underline{k}-points [16$\times$16$\times$1] were used. In
all our calculations, self-consistency has been achieved with a
$0.0001$ eV convergence in total energy. For optimizing the ground
state geometry, forces on the atoms were
converged to less than $0.001$ eV/{\AA} via conjugate gradient
minimization\cite{r30}.\\
\begin{figure}[tb!]
\begin{center}
\includegraphics[scale=0.4]{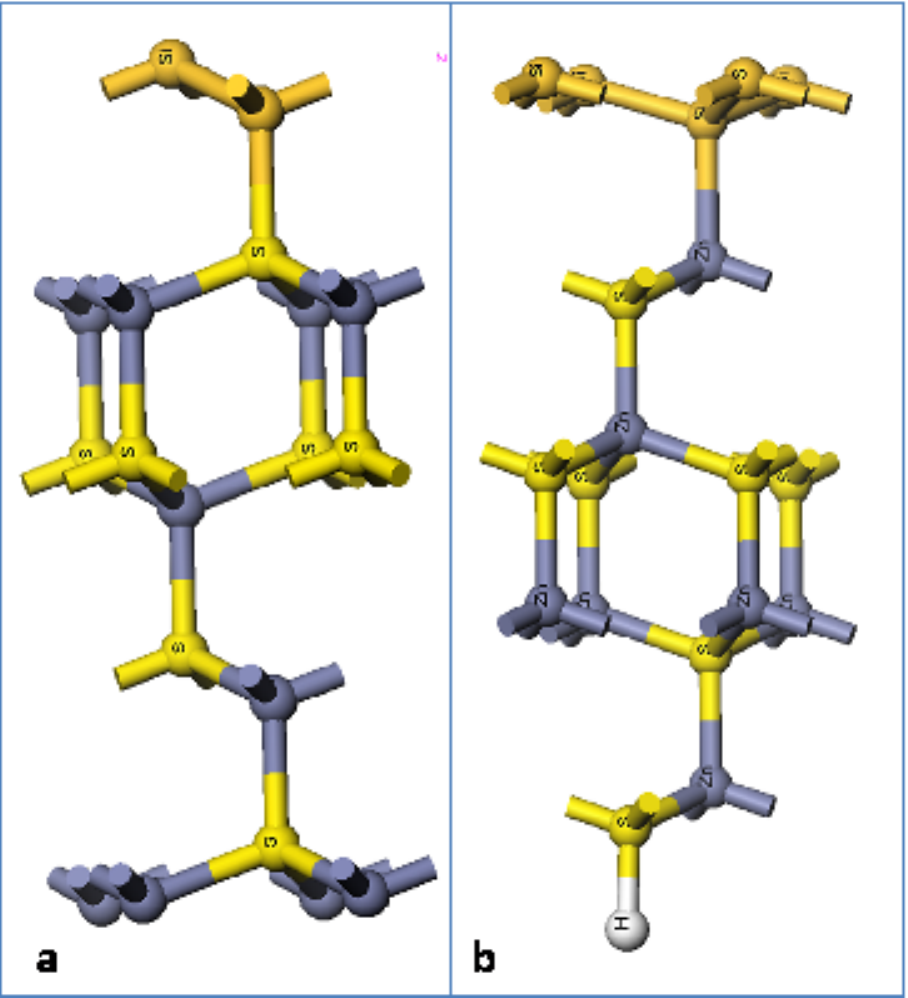}
\caption{Lateral view of one unit cell of epitaxial silicene on semiconductor substrate with Si atom attached to (a) S (non-metal terminated) (b) Zn (metal terminated) surface of ZnS (111). The blue and yellow balls corresponds to the Zn and S atoms of the substrate, while the top overlayer is composed of silicene atoms.}
\label{MTNMT}
\end{center}
\end{figure}
The silicene unit cell has a buckled hexagonal planar geometry,
with a lattice parameter of 3.84 {\AA} and a homogeneous Si-Si bond
distance of 2.29 {\AA} throughout the sheet. The semiconductor
substrates viz. AlAs(111), AlP(111), GaAs(111), GaP(111), ZnS(111) and
ZnSe(111), that we have considered for epitaxial growth
of silicene monolayer, have resonably good lattice matching with that of
silicene, as given in TABLE-\ref{BE}. In this work, we have only considered 
the direct lattice matched non polar-(111) substrates. However, other substrate
orientations may also be relevant and will be addressed in our future study. \\
\begin{table}[tb!]
\caption{Binding energy (in eV/atom) of Silicene monolayer to 
different semiconductor substrates having lattice constants close to that of silicene (3.84\AA).}
\label{BE}
\centering
\begin{tabular}{{p{2.0cm}p{1.8cm}p{1.5cm}p{1.5cm}}}
\hline
Surface & Lat. const.({\AA}) & BE Si@MT
 & BE Si@NMT \\
\hline
AlAs(111)&  4.0038 & 0.630 & 0.370 \\
AlP(111) &  3.8544 & 0.685 & 0.910 \\
GaAs(111)&  3.9974 & 0.475 & 0.880 \\
ZnS(111) &  3.8249 & 0.570 & 0.675 \\
GaP(111) &  3.8541 & 0.605 & 0.525 \\
ZnSe(111)&  4.0076 & 0.440 & 0.540 \\
\hline
\end{tabular}
\end{table}
Therefore, the unit cell of the
composite system consists of the silicene monolayer (having two Si
atoms) placed on top of six layers of the substrate (having 12 atoms), 
with a vacuum slab of 20{\AA} above it. 
\begin{table*}[tb!]
\caption{Magnetic Moment/ unit ($\mu_B$) of the substrate (4$\times$4), before and after introduction of Silicene monolayer on it. The subscript denotes the number of units in the substrate.}
\label{MAG}
\begin{center}
\begin{tabular}{p{1.6cm}p{1.0cm}p{2.0cm}p{1.1cm}p{2.0cm}p{1.1cm}p{2.4cm}p{1.1cm}}
\hline 
NMT  & Mag. Mom. & Si2@NMT  & Mag. Mom. & MT  & Mag. Mom. & Si2@MT  & Mag. Mom. \\
\hline
(As6Al6)$_4$ & 1.41 &  (Si2As6Al6)$_4$&  1.45 & (Al6As6H)$_4$& 0.84 & (Si2Al6As6H)$_4$& 0.00 \\
(P6Al6)$_4$  & 1.42 &  (Si2P6Al6)$_4$ &  1.48 & (Al6P6H)$_4$ & 0.80 & (Si2Al6P6H)$_4$ & 0.00 \\
(As6Ga6)$_4$ & 0.75 &  (Si2As6Ga6)$_4$&  1.26 & (Ga6As6H)$_4$& 0.65 & (Si2Ga6As6H)$_4$ & 0.00 \\
(P6Ga6)$_4$ & 1.39 &  (Si2P6Ga6)$_4$  &  1.49 & (Ga6P6H)$_4$ & 0.78 & (Si2Ga6P6H)$_4$ & 0.86 \\
(Se6Zn6)$_4$ & 0.00 &  (Si2Se6Zn6)$_4$&  0.64 & (Zn6Se6H)$_4$& 0.00 & (Si2Zn6Se6H)$_4$ & 0.38 \\
(S6Zn6)$_4$  & 0.00 &  (Si2S6Zn6)$_4$ &  0.65 & (Zn6S6H)$_4$ & 0.00 & (Si2Zn6S6H)$_4$ & 0.00 \\
\hline
\end{tabular}
\end{center}
\end{table*}
It is to be noted here that in these substrates, the contact layer of the substrate that interacts
with the  silicene overlayer could either be non metal terminated (NMT) or metal
terminated (MT), e.g. in ZnS(111), the interacting layer could be either
S terminated or Zn terminated (FIG.\ref{MTNMT}). If the contact layer is NMT, then the MT
bottom layer in the unit cell does not need H-passivation (FIG.\ref{MTNMT}a). However, 
if the contact is MT, then the
lowest layer of the six layer substrate in the unit cell is NMT that
needs to be passivated by hydrogen (FIG.\ref{MTNMT}b), in order to avoid the effect of
dangling bonds. The nomenclature used for designating an unit of NMT
substrate is NM6M6, while the MT substrate is denoted as M6NM6H. For calculating the
binding energy (BE) and magnetic interaction of the silicene monolayer with the semiconductor substrates
we have used a 4$\times$4 supercell.
The BE of silicene monolayer to the MT and NMT surfaces of these
substrates are summarized in TABLE-\ref{BE}. For MT surface the BE values
lie in the range of 0.56 $\pm$ 0.12
eV/atom (TABLE-\ref{BE}) which is comparable to that of silicene monolayer 
on Ag(111) substrate, for which our estimated value (0.52 eV/atom) is 
in good agreement with first principle result of P. Vogt \textit{et.al} \cite{r17}. 
Thus, on comparing the binding energies of silicene to semiconductor MT/ NMT substrates with Ag(111),
we find that they are close or comparable
for metal terminated surfaces of GaAs(111), GaP(111), ZnS(111), ZnSe(111) substrates and non metal terminated surfaces of AlAs(111), GaP(111) and ZnSe(111) substrates (TABLE-\ref{BE}).\\
Our calculations show that when the bulk substrate (both MT and NMT)
is cleaved to certain number of layers, the top layer contains
uncompensated dangling bonds and therefore, give rise to magnetic
moment in the substrate (TABLE-\ref{MAG}). We have estimated the magnetic moment of the
substrate, before and after introduction of silicene monolayer which are given in TABLE-\ref{MAG}. We would first discuss the behaviour of
silicene with the NMT semiconductor substrates. When silicene monolayer is
introduced to the NMT surface of these substrates, the magnetic moment of the composite
system gets enhanced. For the sake of simplicity, this magnetism can be 
thought upon as the effect of uncompensated dangling bonds at the surface of the 
substrate. However, the 
nature of charge transfer between the silicene layer and substrate may play a vital role in 
enhancement of magnetism of the composite system in all NMT cases. We
have performed Bader charge analysis of NMT substrates before and after
inclusion of silicene layer. In each case, the charge gets transferred from the silicene sheet
to the contact atoms (viz. As, P, S, Se, Ge) of the NMT substrates 
(see TABLE-\ref{bader}), because of the later having higher
electronegavity as compared to Si. Therefore, in all NMT cases the silicene overlayer undergoes 
p-type doping due to its interaction with the substrate.\\
\begin{table}[hb!]
\caption{Bader Charge Analysis of Silicon atoms of the silicene overlayer on MT and NMT substrates.
+ve denotes electron lost while -ve denotes electron gained.}
\begin{center}
\begin{tabular}{lll|lll}
\hline
\multicolumn{3}{p{3.0cm}}{MT}&\multicolumn{3}{p{3.0cm}}{NMT}\\
\hline
System  & Atom & Charge  & System  & Atom & Charge \\
\hline
Si2Al6As6H & Si1 & -0.0475 & Si2As6Al6 & Si1 & +0.001\\
           & Si2 & -0.17   &           & Si2 & +0.33 \\           
Si2Al6P6H & Si1  & -0.05   & Si2P6Al6  & Si1 &+0.0375\\
          & Si2  & -0.16   &           &Si2 & +0.225 \\
Si2Ga6As6H& Si1  & -0.005  & Si2As6Ga6 &Si1 & +0.015 \\
          & Si2  & -0.255  &           &Si2 & +0.2475 \\
Si2Ga6P6H & Si1  & +0.01   & Si2P6Ga6  & Si1 &+0.055 \\
          & Si2  & +0.225  &           &Si2 & +0.2125\\
Si2Zn6Se6H& Si1  & +0.005  & Si2Se6Zn6& Si1 & +0.0275\\
          & Si2  & -0.015  &          &Si2 & +0.0575 \\
Si2Zn6S6H & Si1  & +0.2175 & Si2S6Zn6 & Si1 & +0.425\\
          & Si2  & -0.24   &            &Si2 & -0.2375\\          
\hline
\end{tabular}
\label{bader}
\end{center}
\end{table}
In order to understand the interaction of silicene with these substrates and to help their characterization in the laboratory it is necessary to study the magnetic behavior of silicene with
substrate. Therefore, we have plotted the layer projected DOS and the fattening of silicon bands near the Fermi level in order to study the interaction of the sheet with these substrates. Unlike the NMT cases, the behaviour of silicene monolayer on the MT surface can be 
divided into three categories. The introduction of silicene monolayer 
on MT surface, may quench/ 
enhance the magnetism or may not have any effect at all, on the magnetic moment of the composite (silicene + substrate) system as shown in TABLE-\ref{MAG}. In case 
of AlAs(111), AlP(111) and GaAs(111) [Case-I], the surface magnetism of the MT surface
of the substrate is
quenched after introduction of silicene monolayer and as a result of this, the magnetic moment 
of the composite system is found to vanish. 
Whereas, in GaP(111) and ZnSe(111) [Case-II], the introduction
of silicene monolayer enhances the magnetic moment of the composite system.
However, only in case of ZnS(111) substrate [Case-III], 
the magnetic moment of the MT surface remains `zero' before as well as after 
inclusion of the silicene monolayer.\\ 
The behaviour of silicene on MT surface of semiconductor substrates 
can also be analyzed from the layer projected DOS plots of silicene monolayer on these 
substrates. The LP-DOS plot of silicene monolayer on MT surface show
that it can be metallic or magnetic behaviour, depending
upon the choice of substrate. In FIG.\ref{FATBAND}, we have shown the LP-DOS and band dispersion plots  
of some representative cases [viz. silicene on GaP(111), GaAs(111), ZnS(111) and ZnSe(111)], 
while for the same for AlP(111) and AlAs(111) are given as supplementary information [SI-FIG.1]. 
Silicene on MT surface of AlAs(111),
AlP(111), GaAs(111) and ZnS(111) show metallic behavior with finite density of states
at the Fermi level. FIG.\ref{FATBAND}(b) and (c) show the LP-DOS and band dispersion plots
of silicene on Ga terminated GaAs(111) and Zn terminated ZnS(111) substrate respectively. The 
LP-DOS of silicene on these substrates
resemble the overall DOS of free standing silicene monolayer with
their Dirac cone shifted below the Fermi level by 0.7 eV approximately. 
Silicene monolayer on MT surface of GaP(111), 
show magnetic behavior, with sharp spin up and spin down
splitting near the Fermi level. The spin polarized LP-DOS and band dispersion plots
of silicene on Ga terminated GaP(111) substrate is shown in FIG.\ref{FATBAND}(a). 
It is to be noted here that in case of MT III-V substrates, the silicene bands 
comprises the valence band and conduction bands near the Fermi level. Therefore,
the behaviour of the composite system is dictated by Silicene itself. 
 The spin polarized LP-DOS plot of silicene monolayer on the MT surface of ZnSe(111) show half metallic behavior, with the spin down channel completely vanishing at the Fermi level FIG.\ref{FATBAND}(d). However, in the band dispersion plot of the silicene-substrate composite system for substrates involving Zn [ie. ZnSe(111) and ZnS(111)], it is the Zn $\alpha$-band predominantly constitutes the conduction band bottom and it protrudes below the Fermi level at $\Gamma$ point.\\
 \begin{figure*}[tb!]
\begin{center}
\includegraphics[scale=1.2]{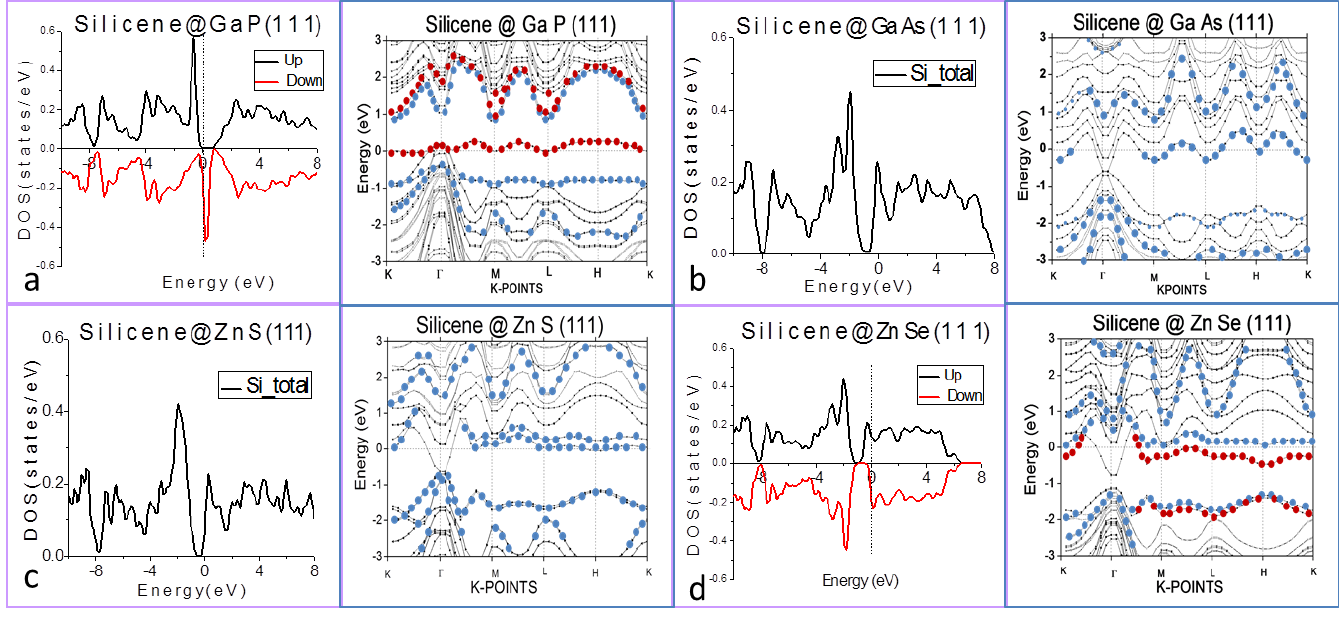}
\caption{Layer projected DOS plot and electronic band dispersion of 
Silicence on; (a):  Ga terminated surface of GaP(111). The blue and red dotted bands 
corresponds to the spin up and spin down components of Si respectively.
(b): Ga terminated surface of GaAs(111) substrate. The blue dotted bands corresponds to those of Silicene. 
(c): Zn terminated surface of ZnS(111) substrate. 
(d): Zn terminated surface of ZnSe(111) substrate. The spin down band (blue dotted)
corresponds to the conduction band only while the spin up band (red dotted) has contribution
in both the valence band and conduction band at the Fermi level giving rise to half 
metallic behaviour.} 
\label{FATBAND}
\end{center}
\end{figure*}
Results of the Bader charge analysis of Silicene monolayer on the MT semiconductor substrates is shown in TABLE-\ref{bader}. In case of most of the MT substrates, electrons are transferred from the substrate to the silicene overlayer because of higher electronegativity of the Si atoms as compared to the contact metal atoms of the substrate. As a result of this the Silicene overlayer undergoes a substrate induced n-type doping\cite{r32,r33}.
While only in case of MT GaP(111) substrate, electrons are transferred from silicene to the substrate [ie. in the opposite direction] which suggest a
p-type doping in the silicene overlayer similar to NMT cases \cite{r34,r35}.\\
\begin{figure}[t!]
\begin{center}
\includegraphics[scale=0.35]{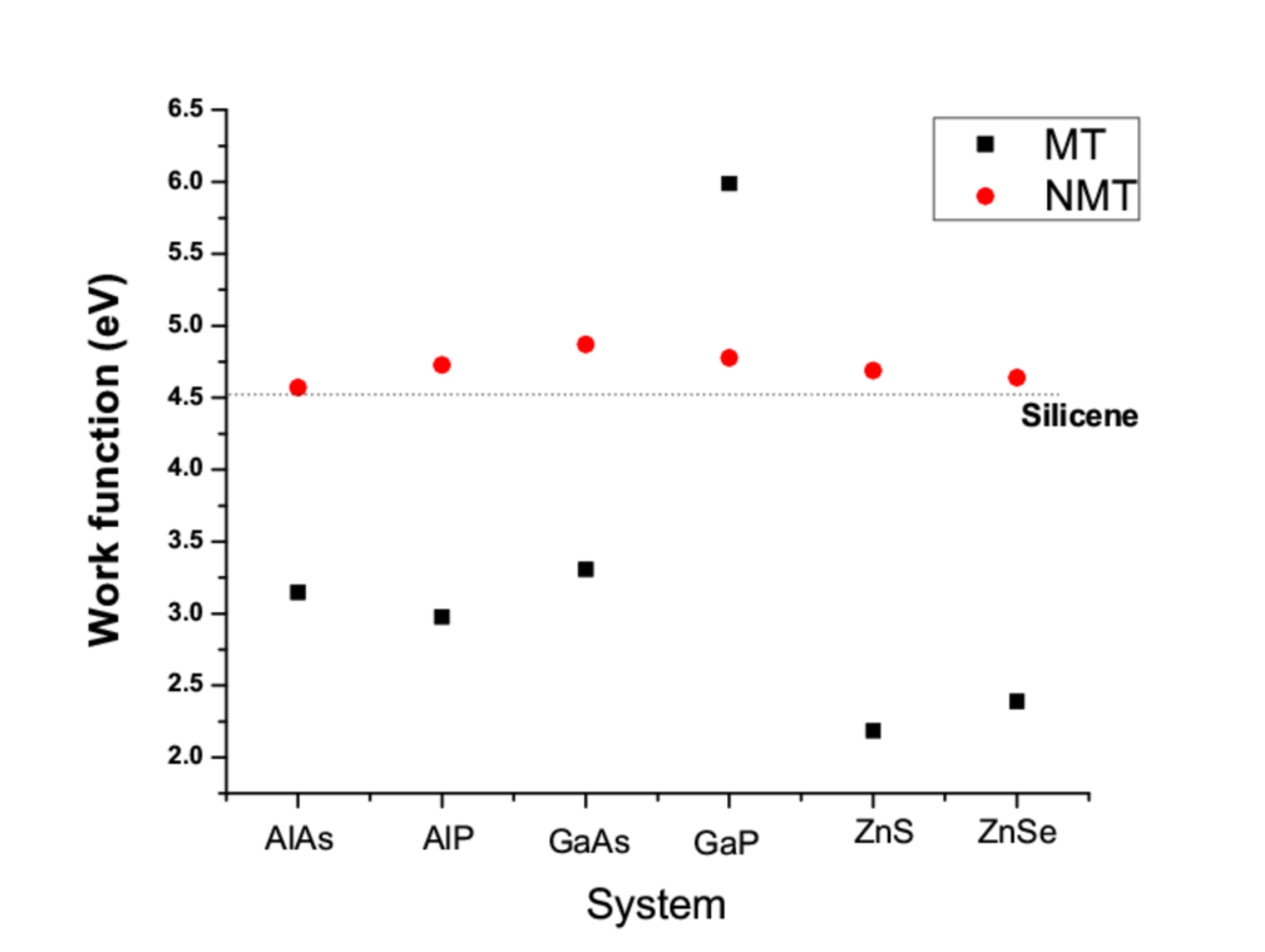}
\caption{Comparison of work function of MT(111) and NMT(111) semiconductor substrates. The dotted line corresponds to the work function of freestanding silicene monolayer} 
\label{WORKFUNCTION}
\end{center}
\end{figure}
For a better understanding of the p-type/ n-type doping of the silicene overlayer, we have estimated the work function (WF) of MT/NMT ($\phi_s$) surfaces of semiconductor substrates and compared it with that of free standing 
silicene monolayer ($\phi_{fl}$). For $\phi_s$ $>$ $\phi_{fl}$, electrons are expected to transfer from silicene overlayer to the substrate, thereby leading to a p-type doping of the silicene overlayer. On the contrary, if $\phi_{fl}$ $>$ $\phi_s$, then electrons should be transferred from the substrate to the silicene overlayer leading to n-type doping. The results of our calculations shown in FIG.\ref{WORKFUNCTION}, reveals that for
MT substrates the WFs are consistently lower than that of free standing silicene (4.5 eV) with the exception of MT GaP(111) substrate (which is reverse to the other MT cases as also seen from the Bader analysis of TABLE-\ref{bader}). However, for NMT surfaces the trend is just reverse leading to the p-type doping. This is just reverse leading to the p-type doping.\\
From the layer projected DOS plot of FIG.\ref{FATBAND}, the difference between the substrate induced p-type and n-type doping of the silicene overlayer can be observed.
In case of GaP(111), the p-type doping of the silicene overlayer is accompanied by shift in Fermi level towards the valence band (FIG.\ref{FATBAND}a) while in the case of n-type doping (FIG.\ref{FATBAND}b, c and d) the Fermi level shifts towards the conduction band. This type of p-type or n-type doping in silicene-substrate composite system could be useful for band gap engineering of bilayer silicene following the case of bilayer graphene \cite{r36,r37}.\\
In conclusion, we report from our first principles based calculations, the
bonding, stability and electronic structure of silicene monolayer,
when epitaxially grown on various Group II-VI, Group III-V and Group IV 
semiconductor substrates viz.
AlAs(111), AlP(111), GaAs(111), GaP(111), ZnS(111) and ZnSe(111). We find that the relative stability and other 
properties of the silicene overlayer depends sensitively on whether
the interacting top layer of the substrate is metal or non-metal terminated. 
The binding energy of silicene monolayer to the metal terminated
surfaces of these substrates are estimated to be range in the 0.56 $\pm$ 0.12 eV/atom. 
The introduction of silicene monolayer on the NMT surface of
 all these semiconductor substrates, leads to enhancement in the magnetic moment
 of the composite system.
However, the behavior of silicene on MT surface of semiconductor substrates can be metallic,
magnetic or semi-metallic depending on the choice of substrate. It undergoes
 substrate induced p-type doping on NMT substrates while n-type doping on MT substrates [with an exception of GaP(111)] depending upon its charge transfer with the substrates.\\
 $^\dagger$ presently affiliated to Fritz-Haber-Institut der Max Planck Gesellschaft, Faradayweg 4-6, Berlin-14195, Germany.\\
$^*$Corresponding Author Email id: msgpd@iacs.res.in

\end{document}